\documentclass[preprint]{elsarticle}
\usepackage{amsfonts}
\usepackage{amsmath}
\usepackage{amssymb}
\usepackage{graphicx}
\usepackage{color}
\begin{document}
\title{Thermalization of electron-positron plasma with quantum degeneracy}
\author[M,U,QO]{M. A. Prakapenia\corref{cor1}}
\ead{nikprokopenya@gmail.com}

\author[M,LTF]{I. A. Siutsou}
\ead{siutsou@icranet.org}

\author[M,P]{G. V. Vereshchagin}
\ead{veresh@icra.it}

\cortext[cor1]{Corresponding author}

\address[M]{ICRANet-Minsk, Institute of physics, National academy of sciences of Belarus\\ 220072 Nezale\v znasci Av. 68-2, Minsk, Belarus}
\address[U]{Department of Theoretical Physics and Astrophysics, Belarusian State University\\ 220030 Nezale\v znasci Av. 4, Minsk, Belarus}
\address[P]{ICRANet, 65122 Piazza della Repubblica, 10, Pescara, Italia}

\begin{abstract}
The non-equilibrium electron-positron-photon plasma thermalization process is studied using relativistic Boltzmann solver, taking into account quantum corrections both in non-relativistic and relativistic cases. Collision integrals are computed from exact QED matrix elements for all binary and triple interactions in the plasma. It is shown that in non-relativistic case (temperatures $k_B T\leq 0.3 m_e c^2$) binary interaction rates dominate over triple ones, resulting in establishment of the kinetic equilibrium prior to final relaxation towards the thermal equilibrium, in agreement with the previous studies. On the contrary, in relativistic case (final temperatures $k_B T\geq 0.3 m_e c^2$) triple interaction rates are fast enough to prevent the establishment of kinetic equilibrium. It is shown that thermalization process strongly depends on quantum degeneracy in initial state, but does not depend on plasma composition.
\end{abstract}


\maketitle

Relativistic plasma is a matter of interest in many fields of physics and astrophysics \cite{2017rkt..book.....V}. One of the peculiar properties of such a plasma is that the mean energy per particle exceeds the rest mass of the electron, so it contains pairs of particles (electrons) and antiparticles (positrons). Attempts to create such a plasma in laboratory \cite{2010PhR...487....1R} are linked with the development of ultra-intense lasers \cite{2012RvMP...84.1177D,2012PhRvL.108p5006R,2014EPJST.223.1083N,2015NatCo...6E6747S,2018NatSR...8.2329E} within large projects, such as ELI \footnote{https://eli-laser.eu/} and XCELS \footnote{http://www.xcels.iapras.ru/}. A number of interesting phenomena such as relativistic transparency \cite{2012NatPh...8..763P}, ultrafast thermalization in magnetized plasma \cite{2015PhRvL.115u5003L} and current-driven instability \cite{2017PhRvL.119r5002W} are predicted and observed in such a plasma. The presence of the pairs greatly enhances plasma opacity, in many cases making it optically thick to photons. In astrophysics, optically thick electron-positron plasma is considered as the source of emission in gamma-ray bursts, soft gamma repeaters, neutron and quark stars and, possibly, blazars. Given highly variable energy release in these sources and their non-thermal spectra the question arises: how such plasma thermalize and how thermalization timescale can be determined.

With the aim of answering this question, relativistic Boltzmann equations for non equilibrium plasma must be solved. In a series of works \cite{2007PhRvL..99l5003A,2008AIPC..966..191A,2008AIPC.1000..309A,2009AIPC.1111..344A} numerical method was developed for a solution of the system of such equations, describing time evolution of distribution functions for electrons, positrons and photons, also in presence protons admixture \cite{2009PhRvD..79d3008A,2010AIPC.1205...11A}. It was shown that not only direct and inverse binary reactions between particles are essential for relaxation to thermal equilibrium, but also triple ones. Thermalization timescales were determined as functions of the total energy density and the baryonic load \cite{2010PhRvE..81d6401A}. Two assumptions were adopted in all these works: 1) quantum degeneracy was neglected and 2) reaction rates for triple processes were computed assuming that detailed balance in all binary reactions (kinetic equilibrium) is established in advance.

In this work we report the study of the relaxation process of uniform isotropic electron-positron-photon plasma towards thermal equilibrium, both in non-relativistic and relativistic cases, relaxing both above mentioned assumptions. The computational scheme is presented in Refs. \cite{2017rkt..book.....V,Siutsou2013,2015AIPC.1693g0007S}. \textcolor{black}{The general scheme of collision integral calculation is presented in \cite{PRAKAPENIA2018533}. Generalization of the scheme to triple interactions is straightforward, details will be given elsewhere.} Time evolution of one-particle distribution functions of electrons $e^-$, positrons $e^+$ and photons $\gamma$ is obtained by numerical integration of relativistic Boltzmann equations \cite{2017rkt..book.....V} including quantum corrections:
\begin{equation}
\frac{1}{c}\frac{\partial f_{i}}{\partial t}=\sum_{q}\left(  \eta_{i}^{q}-\chi_{i}^{q}f_{i}\right),\label{BE}
\end{equation}
where $f_{i}(\epsilon,t)$ are their distribution functions, index $i$
denotes the sort of particles, $\epsilon$ is their energy, $\eta_{i}^{q}$
and $\chi_{i}^{q}$ are the emission and the absorption coefficients of a particle of type "$i$" via the physical process labelled by $q$, $c$ is the speed of light.
The emission and absorption coefficients for the particle $I$ in a binary process $I+II\leftrightarrows III+IV$ have the following form:
\begin{eqnarray}
\label{eta2p}
\eta_{I}^{2p} & = & \int d^3p_2 d^3p_3 d^3p_4 \ W_{(3,4|1,2)} \ f_{III} f_{IV} \left(1+\xi f_{I}\right) \left(1+\xi f_{II}\right), \\
\chi_{I}^{2p}f_{I} & = & \int  d^3p_2 d^3p_3 d^3p_4 \ W_{(1,2|3,4)} \ f_{I} f_{II} \left(1+\xi f_{III}\right) \left(1+\xi f_{IV}\right),
\label{chi2p}
\end{eqnarray}
where transition rates are $W_{(3,4|1,2)}d^3p_3d^3p_4=V dw_{(3,4|1,2)}$ and $W_{(1,2|3,4)}d^3p_1d^3p_2=V dw_{(1,2|3,4)}$, $V$ is normalization volume, $dw$ is differential reaction probability per unit time, $\xi=\psi h^3/2$ and $\psi$ is +1,-1,0 for Bose-Einstein, Fermi-Dirac, Maxwell-Boltzmann statistic, respectively. In what follows we refer to these cases as quantum ($\psi=\pm1$) and classical ($\psi=0$), respectively, $h$ is Planck's constant.

The emission and absorption coefficients for the particle $I$ in a triple process $I+II\leftrightarrows III+IV+V$ have the following form:
\begin{eqnarray}
\eta_{I}^{3p}  = \int  d^3p_2 d^3p_3 d^3p_4 d^3p_5 \ W_{(3,4,5|1,2)} \ f_{III} f_{IV} f_{V} \left( 1+\xi f_{I}\right) \left( 1+\xi f_{II}\right), \\
\label{eta3p}
\chi_{I}^{3p}f_{I} = \int  d^3p_2 d^3p_3 d^3p_4 d^3p_5 \ W_{(1,2|3,4,5)} \ f_{I}f_{II}\left(1+\xi f_{III} \right)\left(1+\xi f_{IV} \right)\left(1+\xi f_{V} \right),
\label{chi3p}
\end{eqnarray}
where $W_{(3,4,5|1,2)}d^3p_3d^3p_4d^3p_5=V dw_{(3,4,5|1,2)}$ and $W_{(1,2|3,4,5)}d^3p_1d^3p_2=V^2 dw_{(1,2|3,4,5)}$.
The expression for $dw$ is given in QED as:
\begin{multline}
dw=c(2\pi \hbar)^4 \delta(\epsilon_{in}-\epsilon_{f\!in})\delta(\mathbf{p}_{in}-\mathbf{p}_{f\!in})|M_{fi}|^2 V \\
\times \left( \prod_{in} \frac{\hbar c}{2\epsilon_{in} V}\right) \left(  \prod_{f\!in} \frac{d^3p_{f\!in}}{(2\pi\hbar)^3} \frac{\hbar c}{2\epsilon_{f\!in}}\right),
\end{multline}
where $\mathbf{p}_{f\!in}$ and $\epsilon_{f\!in}$ are respectively
momenta and energies of outgoing particles, $\mathbf{p}_{in}$ and $\epsilon_{in}$ are momenta and energies of
incoming particles, $M_{fi}$ is the corresponding matrix
element, $\delta$-functions stand for energy-momentum conservation, $\hbar=h/2\pi$.
Therefore, collision integrals, i.e. right-hand side of eqs. (\ref{BE}), are integrals over the phase space of interacting particles, which include the QED matrix elements, see e.g. \cite{Berestetskii1982,2017rkt..book.....V} for binary reactions and \cite{1952RSPSA.215..497M} for double Compton scattering, \cite{2004epb..book.....H} for relativistic bremsstrahlung and \cite{1976spr..book.....J} for substitution rules in computation of remaining matrix elements for triple reactions. We consider all binary and triple interactions between electrons, positrons and photons as listed in Tab. \ref{tab1}. 
\begin{table}[tbp] \centering
\begin{tabular}
[c]{|c|c|}\hline
Binary processes & Triple processes\\\hline\hline
{M{\o }ller, Bhabha} & {Bremsstrahlung}\\
{$e^{\pm}{e^{\pm\prime}\leftrightarrow e^{\pm}}^{\prime\prime}$}${e^{\pm}%
}^{\prime\prime\prime}$ & {$e^{\pm}e^{\pm\prime}{\leftrightarrow}e^{\pm
\prime\prime}e^{\pm\prime\prime\prime}\gamma$}\\
{$e^{\pm}{e^{\mp}\leftrightarrow e^{\pm\prime}}$}${e^{\mp}}^{\prime}$ &
{$e^{\pm}e^{\mp}{\leftrightarrow}e^{\pm\prime}e{^{\mp\prime}}\gamma$}\\\hline
Single {Compton} & {Double Compton}\\
{ $e^{\pm}\gamma{\leftrightarrow}e^{\pm}\gamma^{\prime}$} & {$e^{\pm}%
\gamma{\leftrightarrow}e^{\pm\prime}\gamma^{\prime}\gamma^{\prime\prime}$%
}\\\hline
{Pair production} & Radiative pair production\\
and annihilation & and three photon annihilation\\
{$\gamma\gamma^{\prime}{\leftrightarrow}e^{\pm}e^{\mp}$} & $\gamma
\gamma^{\prime}${${\leftrightarrow}e^{\pm}e^{\mp}$}$\gamma^{\prime\prime}$\\
& $e^{\pm}\gamma${${\leftrightarrow}e^{\pm\prime}{e^{\mp}}e^{\pm\prime\prime}%
$}\\
& {$e^{\pm}e^{\mp}{\leftrightarrow}\gamma\gamma^{\prime}$}$\gamma
^{\prime\prime}$\\\hline
\end{tabular}
\caption{Binary and triple QED processes in the pair plasma.}\label{tab1}%
\end{table}
The coupled system of integro-differential equations (\ref{BE}) is solved numerically using a finite difference method by introducing a computational grid in the phase space to represent the distribution functions and to compute collisional integrals. \textcolor{black}{We introduce spherical coordinates in momentum space, they are energy and two angles for momentum direction. We use logarithmic energy grid with 60 nodes both for binary and triple interactions. Angular grid contains 64x128 nodes for binary interactions and 24x48 nodes for triple interactions. Computational time for collisional integrals of triple interactions is much greater than that of binary interactions (about 1 week vs. 1 hour).} As particle interactions have different rates (timescales) this system of equations is stiff: it is solved with implicit Gear method \cite{1976oup..book.....H}.

The kinetic equilibrium is a state of plasma in which all direct and inverse binary reactions compensate each other $\eta_{i}^{2p}=\chi_{i}^{2p}f_{i}$. As a consequence, in such a state, all components have a common temperature and chemical potential, and their distribution functions have a Fermi-Dirac/Bose-Einstein shape:
\begin{equation}
f_{i}(\varepsilon)=\left(\frac{1}{2\pi \hbar}\right)^3\left[ \exp\left(\frac{\varepsilon-\varphi_{i}}{\theta_i}\right) -\psi \right]^{-1},\label{df}
\end{equation}
with chemical potentials $\varphi_i \equiv \mu_i/m_e c^2$ and temperatures $\theta_i \equiv k_B T_i/m_e c^2$, where $\varepsilon \equiv \epsilon/m_e c^2$ is the energy of the particle, $m_e$ is electron mass, $k_B$ is Boltzmann's constant. In this work energy density spectra $d\rho_i/d\epsilon=4\pi |\mathbf{p}|\epsilon^2 c^{-2}f_i$ are used instead of distribution functions.
One can estimate a timescale of emergence of kinetic equilibrium starting from arbitrary non-equilibrium state as $\tau_{2p}^{-1}=(n_++n_-) c\sigma_T$, where $\sigma_T$ is Thompson cross-section.

Thermal equilibrium is a state in which all direct and inverse reactions compensate each other. In such a state, all plasma components obey distributions (\ref{df}) with equilibrium temperature and zero chemical potential. Assuming that cross-sections of triple processes are limited by the value $\alpha \sigma_T$, one can estimate the corresponding timescale as $\tau_{3p}^{-1}=\alpha(n_{+}+n_{-})c\sigma_T$.
\begin{figure}
[ptb!]
\begin{center}
\includegraphics[width=90mm]{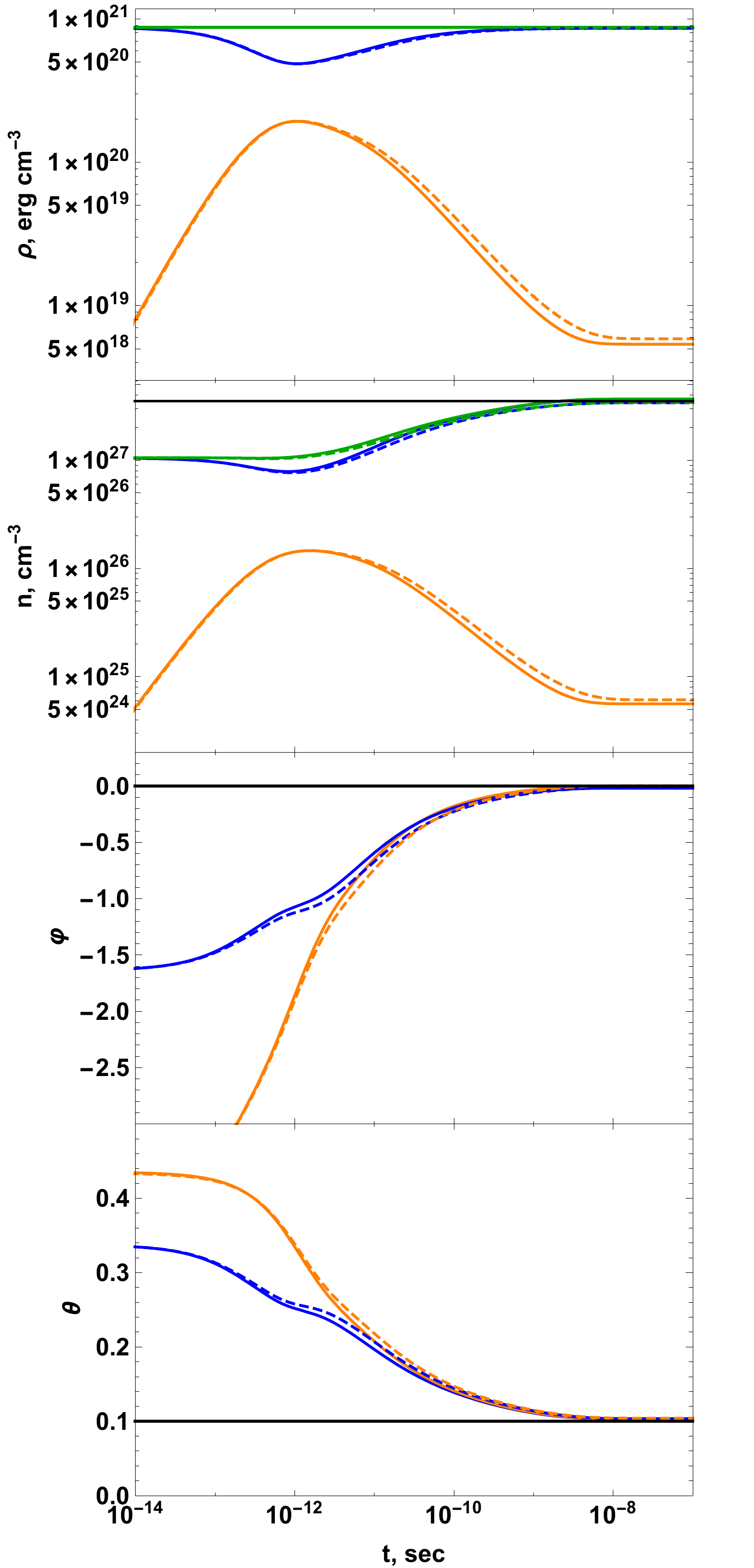}
\caption{From top to bottom: time evolution of energy density, particle number density, chemical potential and temperature of photons (blue), electrons/positrons (orange), all together (green). Black line represents the final equilibrium quantities. Solid (dashed) curves are obtained for quantum (classical, with $\psi=0$) statistics. Final equilibrium temperature is $\theta_{f\!in}=0.1$.}
\label{therm0.1}
\end{center}
\end{figure}
We have performed a number of numerical simulations in a range of energy densities corresponding to the final temperatures $0.1<\theta_{f\!in}<4.5$ with arbitrary initial energy density spectra. We also considered two different initial states: electrons and positrons only (photoless) and photons only (pairless). 
\begin{figure}
[ptb!]
\begin{center}
\includegraphics[width=70mm]{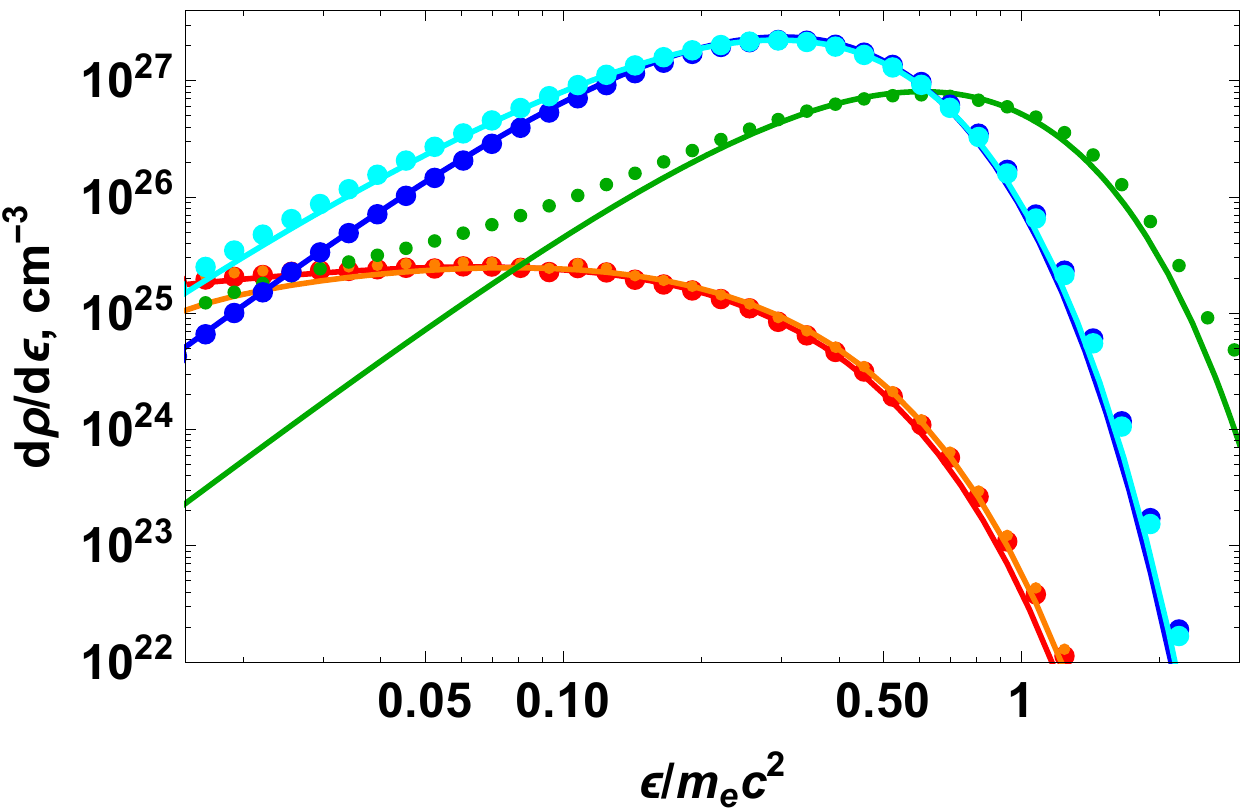}
\caption{Energy density spectra at selected time moments (dots). Solid curves represent fits to numerical results: photon Boltzmann spectrum (blue), photon Bose-Einstein spectrum (cyan), pairs Boltzmann spectrum (orange), pairs Fermi-Dirac energy spectrum (red). Green solid curve represents photon Bose-Einstein spectrum fit at $t=10^{-11}$ sec with $\varphi_{kin}=-0.65, \theta_{kin}=0.2$. Final equilibrium temperature  is $\theta_{f\!in}=0.1$.}
\label{spfin0.1}
\end{center}
\end{figure}

First, we present a particular result for non-relativistic case with total energy density $\rho_{tot}=8.7\times 10^{20}\,\text{erg cm}^{-3}$ corresponding to a final equilibrium temperature $\theta_{f\!in}=0.1$. Total initial particle number density is $n_{tot}^{in}=0.3 n_{tot}^{f\!in}$, where  $n_{tot}^{f\!in}=3.5\times 10^{27}\,\text{cm}^{-3}$ is the final total particle number density in equilibrium. The initial energy density spectrum has a power law shape $d\rho/d\epsilon\sim \epsilon^{-\kappa}$, where $\kappa$ is a constant. The initial state is pairless. The time evolution of basic thermodynamic quantities is shown on Fig.~\ref{therm0.1}, while energy density spectrum for selected time moments is shown on Fig.~\ref{spfin0.1}. Chemical potentials and temperatures are computed from the total energy and the number densities \cite{2009PhRvD..79d3008A}. Total energy density does not change in time due to energy conservation. Total particle number density changes only due to imbalance in triple processes. As the initial state of plasma is pairless, we observe a growth of a pair number density up to $10^{-13}$ sec. The kinetic equilibrium is established at $t\simeq\tau_{2p}$, when chemical potentials and temperatures of plasma components become equal: at $t=10^{-11}$ sec we find $\varphi_\gamma=\varphi_\pm=-0.65$ and $\theta_\gamma=\theta_\pm=0.2$,  see Fig.~\ref{therm0.1}. At this point the spectrum has the shape (\ref{df})  near its maximum, and deviations occur only at low and high energy tails, which thermalize at later times, compare cyan and green curves on Fig.~\ref{spfin0.1}. From this moment on, chemical potentials and temperatures shown on Fig. \ref{therm0.1} acquire physical meaning. On Fig.~\ref{spfin0.1} we also present the final particle spectra fulfilling quantum and classical, setting $\xi=0$ in equations (\ref{eta2p})-(\ref{chi3p}), statistics. The photon spectra have respectively Planck and Boltzmann shapes: compare Rayleigh-Jeans and Wien low energy slopes in cyan and blue curves.  
The thermal equilibrium is reached with zero chemical potential and final temperature $\theta_{f\!in}=0.1$ at $t\sim10^{-8}$ sec. Analogous evolution is observed for the photonless initial state. Dashed and solid curves on Fig. \ref{therm0.1} coincide up to few percent, so Bose enhancement and Pauli blocking factors in eqs. (\ref{eta2p})-(\ref{chi3p}) are not significant. These results are in full agreement with a previous work \cite{2009PhRvD..79d3008A}. 
\begin{figure}
[ptb!]
\begin{center}
\includegraphics[width=70mm]{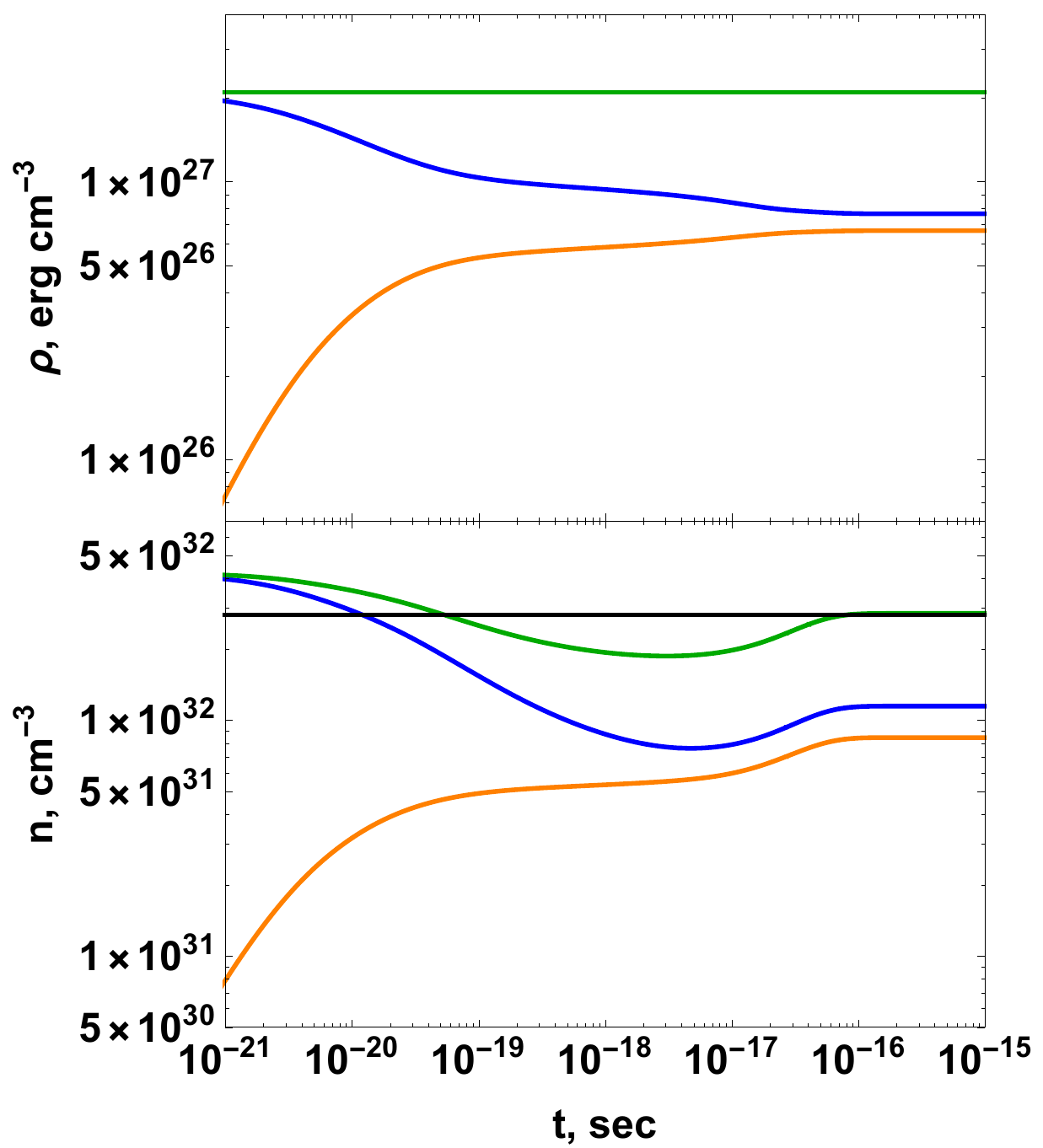}
\caption{Time evolution of energy density and particle number density for relativistic plasma with quantum statistics. The final equilibrium temperature is $\theta_{f\!in}=3$. Colors are as in Fig. \ref{therm0.1}.}
\label{therm3}
\end{center}
\end{figure}

\begin{figure}
[ptb!]
\begin{center}
\includegraphics[width=70mm]{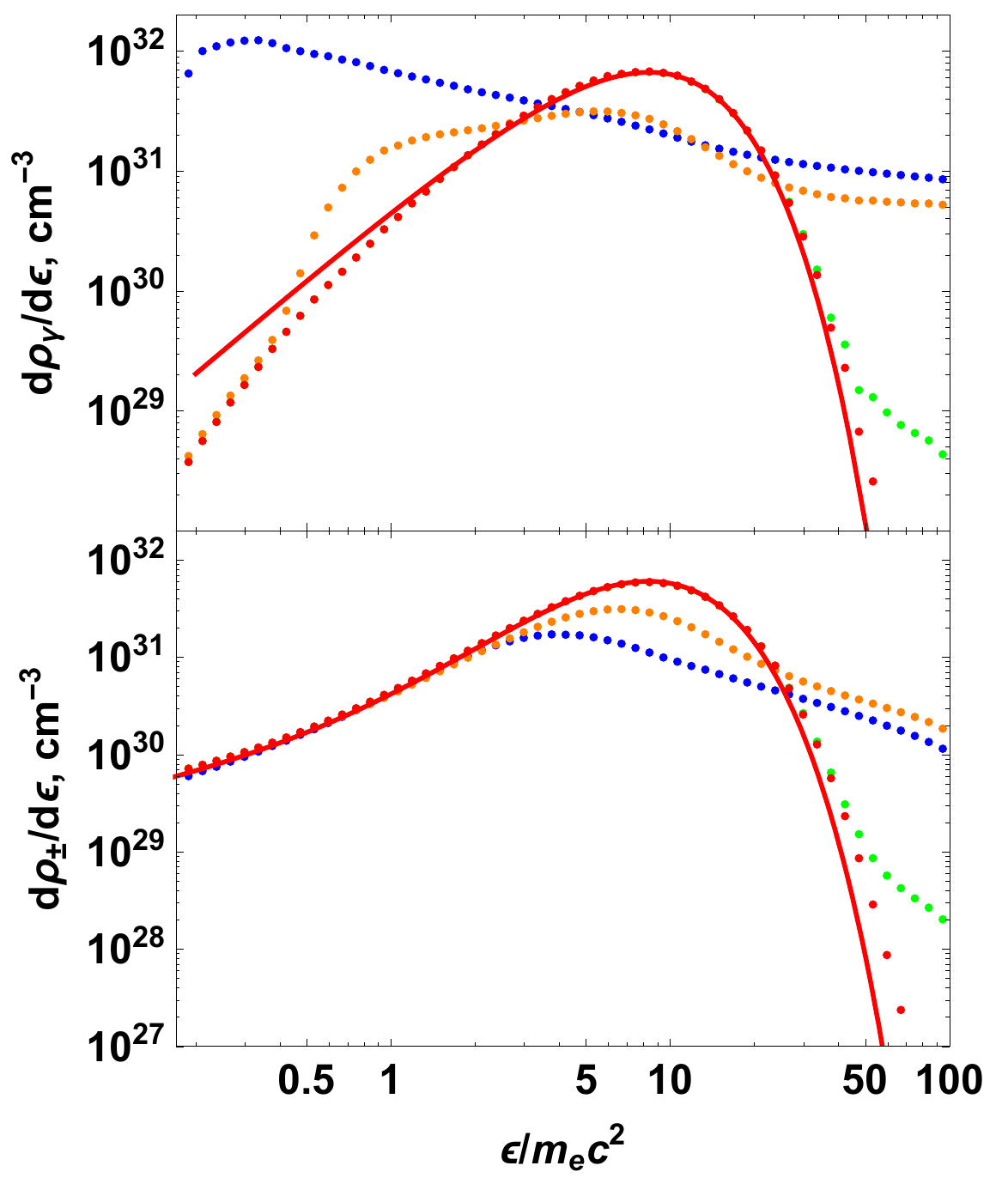}
\caption{Photon and pair quantum energy spectrum at different moments of time evolution: $t=10^{-20}$ sec (blue), $t=10^{-18}$ sec (orange), $t=10^{-16}$ sec (green), $t=10^{-15}$ sec (red). The final equilibrium temperature is $\theta_{f\!in}=3$.}
\label{spev3}
\end{center}
\end{figure}

Second, we present the result for relativistic case with total energy density $\rho_{tot}=2.1\times 10^{27}\,\text{erg cm}^{-3}$ corresponding to a final equilibrium temperature $\theta_{f\!in}=3$. Initial total particle number density is $n_{tot}^{in}=1.5 n_{tot}^{f\!in}$, where  $n_{tot}^{f\!in}=2.8\times 10^{32}\,\text{cm}^{-3}$ is the final total particle number density in equilibrium. The initial state is pairless, with a power law spectrum. Time evolution of energy density and particle number density is shown on Fig.~\ref{therm3}. At the beginning pair number density is increasing with time up to $t=10^{-20}$ sec. Instead of chemical potential and temperature, we show energy density spectra on Fig.~\ref{spev3}. It is clear that spectral evolution starts much earlier in this case, with the low energy part of the spectrum approaching thermal shape as early as at $10^{-18}$ sec. The spectrum acquires the shape (\ref{df}) at $t\simeq \tau_{3p}\simeq 10^{-16}$ sec (green dots), with $\varphi_{fin}=0$ and $\theta_{fin}=3$. Unlike the non-relativistic case, the spectrum at time $\tau_{2p}\simeq 10^{-18}$ sec (orange dots) is far from equilibrium. Such a behaviour of the spectrum shows that the kinetic equilibrium stage is absent and relativistic plasma relaxes directly to thermal equilibrium. Note that deviations at low energy part of the photon spectrum from final equilibrium fit are caused by the loss of numerical precision in calculation of triple reactions, which is of a high computational cost.

\begin{figure}
[ptb!]
\begin{center}
\includegraphics[width=70mm]{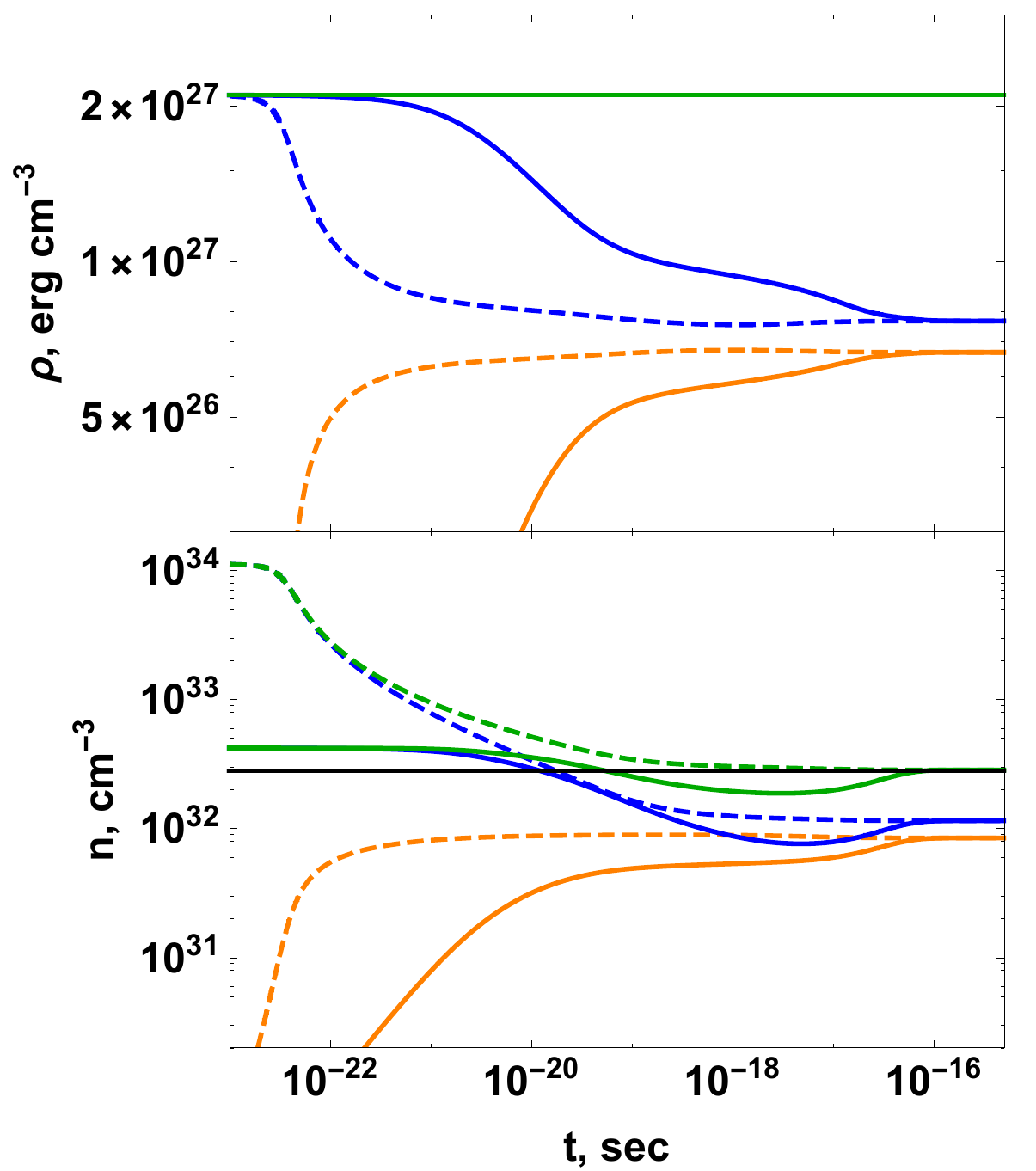}
\caption{Time evolution of energy density and particle number density for relativistic plasma with quantum statistics. Solid and dashed curves correspond to the cases $n^{in}_{tot}=1.5n^{f\!in}_{tot}$ and $n^{in}_{tot}=40n^{f\!in}_{tot}$, respectively. The final equilibrium temperature is $\theta_{fin}=3$. Colors are as in Fig. \ref{therm0.1}.}
\label{thermdeg}
\end{center}
\end{figure}

Third, we discuss the role of Bose enhancement and Pauli blocking terms in eqs. (\ref{eta2p})-(\ref{chi3p}) in highly degenerate plasma. The degree of plasma degeneracy can be parametrized as $D=(n_{tot}\lambda_{th}^3)^{-1}$, where $\lambda_{th}=c \hbar /(k_B T)$ is thermal wavelength. \textcolor{black}{This parameter depends on temperature (or total energy of plasma) and total particle number, which can take arbitrary values.} Degenerate plasma has $D<1$. Note that Pauli principle imposes strong constraint on the choice of initial state, so total electron (or positron) number density in a fully degenerate relativistic plasma cannot exceed $\approx5$ times its thermal value. Therefore we choose pairless initial state. On Fig. \ref{thermdeg} we compare thermalization processes for relativistic plasma with quantum statistics for different initial states with \textcolor{black}{$D=1.12~(n_{tot}=1.5\times2.8\times10^{32} \text{cm}^{-3}, k_B T=3m_ec^2)$ and $D=0.0419~(n_{tot}=40\times2.8\times10^{32} \text{cm}^{-3}, k_B T=3m_ec^2)$.}

Clearly, the thermalization process in these two cases is different. In highly degenerate plasma, electron-positron pairs are created much earlier, at about $10^{-22}$ sec, compared to $10^{-20}$ sec in plasma with $D\sim 1$. This is essentially due to the presence of Bose enhancement terms in eqs. (\ref{eta2p})-(\ref{chi3p}). The dominance of Bose enhancement terms is verified by performing simulations with identical initial conditions neglecting these terms. Despite this obvious difference, the final equilibrium state is reached almost simultaneously in both cases, because of the slower progress to final equilibrium in the highly degenerate case. Indeed, thermalization timescales for low and high degenerate cases are $2.0\times 10^{-17}$ sec and $1.1\times 10^{-17}$ sec, respectively.

\begin{figure}
[ptb!]
\begin{center}
\includegraphics[width=70mm]{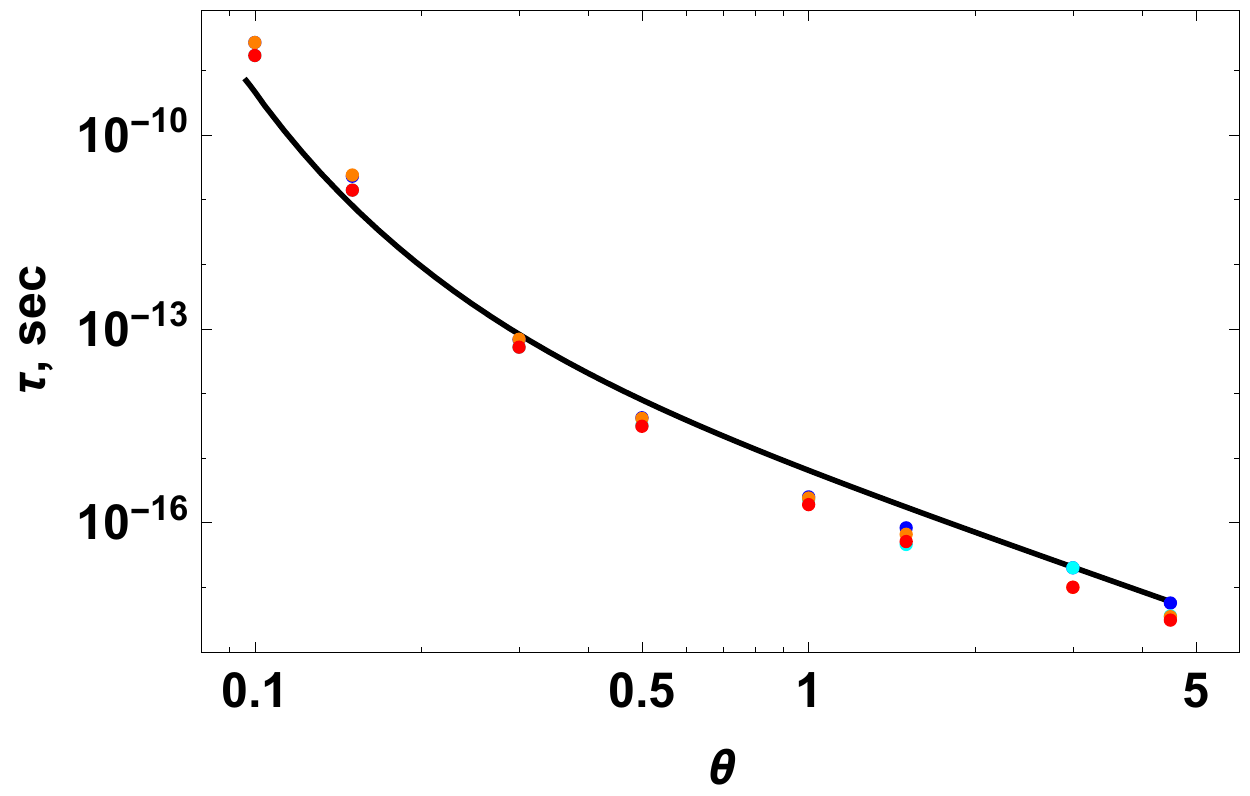}
\caption{Thermalization timescales for electron-positron-photon plasma as a function of the final temperature. Classical statistics with initial pairless state (blue), quantum statistics with initial pairless state (cyan), classical statistics with initial photonless state (orange), quantun statistics with initial photonless state (red). Black curve represents estimated timescales $\tau_{3p}.$}%
\label{timesc}
\end{center}
\end{figure}

One can estimate thermalization timescale $\tau_{th}$ by considering the approach of any thermodynamic variable to its final value \cite{2010PhRvE..81d6401A}. In this work, we fit the photon energy density with a function $\rho_{\gamma}(t)=\rho_{\gamma}^{f\!in}+A\exp(-t/\tau_{th})$, where $A$ is a constant. We solve the optimization problem for different time intervals near the final state, and find best fit to numerical data. Our results are presented on Fig. \ref{timesc}, compared with the timescale $\tau_{3p}$. Surprisingly, such a rough estimate reasonably agrees with our results. Nevertheless, computed timescales may differ from estimates essentially. Inspection of Fig. \ref{timesc} shows that thermalization occurs slightly faster when quantum corrections in collision integrals are taken into account. Photonless and pairless initial states thermalize on similar timescales.

In this work, for the first time, we computed collision integrals of triple interactions by integration of QED matrix elements, in full analogy with binary ones. Our results show that the assumption made in the previous works \cite{2007PhRvL..99l5003A,2009PhRvD..79d3008A}, that reaction rates of triple processes are $\alpha$ times smaller than binary ones, is valid for non-relativistic case. Indeed, detailed balance in binary processes occurs before triple ones, and kinetic equilibrium is set before the final equilibrium is established, see Fig.~\ref{therm0.1}, where total particle number density $n_{tot}$ is almost constant at early times and it starts to change after the establishment of kinetic equilibrium.
We found that reaction rates for triple processes are no longer subdominant, in particular bremsstrahlung and double Compton scattering play crucial roles in thermalization process changing the total number density of particles at sufficiently early times (compared to non-relativistic case). The spectrum at low energies approaches thermal equilibrium form, whereas high energy part of the spectrum relaxes to equilibrium at later times. As a consequence, in relativistic case kinetic equilibrium is absent, both binary and triple processes simultaneously reach detailed balance at the final thermal equilibrium state.
Reaction rates in initially pairless degenerate plasma are enhanced by orders of magnitude, so the role of Bose enhancement in plasma kinetics is crucial at early times. As soon as number densities approach thermal values, plasma degeneracy decreases. In fact, thermalization timescales are determined by relaxation kinetics near equilibrium, and Bose enhancement and Pauli blocking factors at this stage are irrelevant. Finally, performing simulations with photonless degenerate plasma we do not find significant effects of Pauli blocking in thermalization process.

In summary, in this work relaxation of relativistic pair plasma is studied for the first time using relativistic Boltzmann equations with  quantum corrections. Collision integrals for all binary and triple reactions are computed from the first principles. Thermalization timescales in the final temperature range $0.1\leq \theta_{f\!in}\leq 4.5$ are determined. We show that kinetic equilibrium is established prior to the thermal equilibrium in non-relativistic case ($\theta_{f\!in}<0.3$) and is absent in relativistic case ($\theta_{f\!in}>0.3$). Finally, quantum degeneracy leads to enhanced pair production in initially pairless state.

{\bf Acknowledgements.} We would like to thank Dr. Alexey Aksenov for providing us with the earlier versions of the Boltzmann solver.

\bibliographystyle{aip}
\bibliography{pair}

\end{document}